\documentclass[runningheads,a4paper]{llncs}

\usepackage{amsmath}
\usepackage{mathrsfs}
\usepackage{marvosym}
\usepackage{bbm}
\usepackage{etex}
\usepackage{amssymb}
\usepackage{graphicx}
\usepackage{stmaryrd}
\usepackage[all]{xy}
\usepackage{color}
\usepackage{tikz-qtree}
\usepackage{bussproofs}
\usepackage[justification=centering,labelfont=bf]{caption}

\usepackage{url}
\urldef{\mailsa}\path{grellois@pps.univ-paris-diderot.fr}
\urldef{\mailsb}\path{mellies@pps.univ-paris-diderot.fr}
\newcommand{\keywords}[1]{\par\addvspace\baselineskip
\noindent\keywordname\enspace\ignorespaces#1}

\newcommand\red[1]{#1}

\newcommand{\superbang}{\lightning} 
\newcommand{\modality}{\Box}
\newcommand{\colorbang}{\lightning\hspace{-.5765em}\lightning\hspace{-.5765em}\lightning}
\newcommand{\Rel}{Rel}
\newcommand{\Relinfinitary}{\underline{Rel}}
\newcommand{\with}{\&}
\newcommand{\tensor}{\otimes}
\newcommand{\lax}{m}
\newcommand{\laxzero}{m^0}
\newcommand{\laxdeux}[2]{m^2_{#1 , #2}}
\newcommand{\LAT}{\mathscr{L}}
\newcommand{\Ccategory}{\mathscr{C}}
\newcommand{\dig}[1]{\mathbf{dig}_{#1}}
\newcommand{\der}[1]{\mathbf{der}_{#1}}
\newcommand{\diag}[1]{\Delta_{#1}}

\newcommand{\morph}[1]{\stackrel{#1}{\longrightarrow}}
\newcommand{\one}{1}
\newcommand{\paire}[2]{\langle #1,#2\rangle}
\newcommand{\fixrule}{fix}
\newcommand{\runtree}[2]{\textbf{run-tree}(#1,#2)}
\newcommand{\tree}{\textit{witness}}
\newcommand{\leaves}[1]{\textbf{leaves}(#1)}
\newcommand{\negation}[1]{#1^{\bot}}
\newcommand{\fixpoint}[1]{\textbf{Y}_{#1}}
\newcommand{\cardreals}{2^{\aleph_0}}

\begin{document}

\mainmatter  

\title{An infinitary model of linear logic}

\author{Charles Grellois \and Paul-Andr\'e Melli\`es}
\authorrunning{Charles Grellois \and Paul-Andr\'e Melli\`es}

\institute{Universit\'e Paris Diderot, Sorbonne Paris Cit\'e\\
Laboratoire Preuves, Programmes, Syst\`emes\\
\mailsa\\
\mailsb}

\maketitle

\begin{abstract}
In this paper, we construct an infinitary variant of the relational model of linear logic, 
where the exponential modality is interpreted as the set of finite or countable multisets. 
We explain how to interpret in this model the fixpoint operator~$\fixpoint{}$ as a Conway operator 
alternatively defined in an inductive or a coinductive way. We then extend the relational semantics
with a notion of color or priority in the sense of parity games. This extension enables us 
to define a new fixpoint operator~$\fixpoint{}$ combining both inductive and coinductive policies. 
We conclude the paper by mentionning a connection between the resulting model of
$\lambda$-calculus with recursion and higher-order model-checking.
\keywords{Linear logic, relational semantics, fixpoint operators, induction and coinduction,
parity conditions, higher-order model-checking.}
\end{abstract}

\abovedisplayskip = 5pt
\belowdisplayskip = 5pt
\abovedisplayshortskip = 5pt
\belowdisplayshortskip = 5pt
\section{Introduction}

%
%
%
%
In many respects, denotational semantics started in the late 1960's with Dana Scott's introduction of domains and the fundamental intuition that
$\lambda$-terms should be interpreted as \emph{continuous} rather than general functions between domains.
This seminal insight has been so influential in the history of our discipline that it remains 
deeply rooted in the foundations of denotational semantics more than fourty-five years later.
In the case of linear logic, this inclination for continuity means that 
the interpretation of the exponential modality
$$
A \quad \mapsto \quad !\, A
$$
is \emph{finitary} in most denotational semantics of linear logic.
This finitary nature of the exponential modality is tightly connected to continuity
because this modality regulates the linear decomposition of the intuitionistic implication:
$$A \, \Rightarrow \, B \quad = \quad !\, A \, \multimap \, B.$$
%
Typically, in the qualitative and quantitative 
coherence space semantics of linear logic, 
the coherence space $!\,A$ is either defined as the coherence space~$!A$ 
of \emph{finite} cliques (in the qualitative semantics) or of \emph{finite} multi-cliques (in the quantitative semantics)
of the original coherence space~$A$.
This finiteness condition on the cliques $\{a_1,\dots,a_n\}$ or multi-cliques $[a_1,\dots,a_n]$ of the coherence space $!A$ 
captures the computational intuition that, in order to reach a given position~$b$ of the coherence space~$B$,
every proof or program
$$
f \quad : \quad !\, A \, \multimap \, B
$$
will only explore a \emph{finite} number of copies of the hypothesis~$A$, 
and reach at the end of the computation a specific position~$a_i$ in each copy of the coherence space~$A$.
In other words, the finitary nature of the interpretation of $!A$ is just an alternative
and very concrete way to express in these traditional models of linear logic
the continuity of proofs and programs.


In this paper, we would like to revisit this well-established semantic tradition and accomodate another equally well-established tradition,
coming this time from verification and model-checking.
We find especially important to address and to clarify an apparent antagonism between the two traditions.
%
%
Model-checking is generally interested in infinitary (typically $\omega$-regular) inductive and coinductive behaviours 
of programs which lie obviously far beyond the scope of Scott continuity.
For that reason, we introduce a variant of the relational semantics of linear logic 
where the exponential modality, noted in this context
$$
A \quad \mapsto \quad \superbang \, A
$$
is defined as the set of \emph{finite} or \emph{countable} multisets of the set~$A$.
From this follows that a proof or a program
$$
A \, \Rightarrow \, B \quad = \quad \superbang\, A \, \multimap \, B.
$$
%
is allowed in the resulting infinitary semantics to explore
a possibly countable number of copies of his hypothesis~$A$ 
in order to reach a position in~$B$.
By relaxing the continuity principle, this mild alteration of the original relational semantics
paves the way to a fruitful interaction between linear logic and model-checking.
%
%
This link between linear logic and model-checking is supported by the somewhat unexpected observation 
that the binary relation 
$$Y(f) \quad : \quad ! X \quad \morph{} \quad A$$
defining the fixpoint $\fixpoint{}(f)$ associated to a morphism
$$f \quad : \quad ! X \, \tensor \, !A \quad \morph{} \quad A$$
in the familiar (and thus finitary) relational semantics of linear logic 
is defined by performing a series of explorations of the infinite binary tree
$$
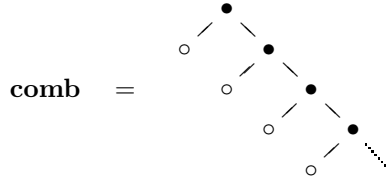
\begin{array}{ccc}
\textbf{comb}
&
\quad = \quad &
\vcenter{\xymatrix @-1.6pc {
&\bullet \ar@{-}[rd]\ar@{-}[ld]
\\
\circ && \bullet \ar@{-}[rd]\ar@{-}[ld]
\\
& \circ && \bullet \ar@{-}[rd]\ar@{-}[ld]
\\
&& \circ && \bullet \ar@{.}[rd]\ar@{-}[ld]
\\
&&& \circ &&
}}
\end{array}
$$
by an alternating tree automaton $\langle \, \Sigma \, , \, Q \, , \, \delta_f \, \rangle$
on the alphabet $\Sigma=\{\bullet,\circ\}$ defined by the binary relation~$f$.
The key idea is to define the set of states of the automaton as $Q=A\uplus X$ and to associate a transition 
$$
\delta_f(\bullet,a) \quad = \quad (\, x_1\wedge \dots \wedge x_k \, , \,  a_1\wedge \dots \wedge a_n \, )
$$

\noindent
of the automaton to any element $(([x_1,\dots,x_k],[a_1,\dots, a_n]),a)$ of the binary relation $f$,
where the $x_i$'s are elements of~$X$ and the $a_i$'s are elements of~$A$~;
and to let the symbol~$\circ$ accept any state $x\in X$.
%
%
Then, it appears that the traditional definition of the fixpoint operator $\fixpoint{}(f)$ as a binary relation $!X\to A$
may be derived from the construction of run-trees of the tree-automaton $\langle \, \Sigma \, , \, Q \, , \, \delta_f \, \rangle$
on the infinitary tree $\textbf{comb}$.
%
More precisely, the binary relation $Y(f)$ contains all the elements $([x_1,\dots,x_k],a)$
such that there exists a finite run-tree (called $\tree$) of the tree automaton $\langle \, \Sigma \, , \, Q \, , \, \delta_f \, \rangle$
accepting the state~$a$ with the multi-set of states $[x_1,\dots,x_k]$ collected at the leaves $\circ$.
As far as we know, this automata-theoretic account of the traditional construction of the fixpoint operator $\fixpoint{}(f)$ 
in the relational semantics of linear logic is a new insight of the present paper, which we carefully develop in~\S\ref{section/finitary-fixpoint}.

\medbreak

Once this healthy bridge between linear logic and tree automata theory identified,
it makes sense to study variations of the relational semantics inspired by verification.
This is precisely the path we follow here by replacing the finitary interpretation $!A$
of the exponential modality by the finite-or-countable one $\superbang A$.
This alteration enables us to define an inductive as well as a coinductive fixpoint operator $\fixpoint{}$
in the resulting infinitary relational semantics.
The two fixpoint operators only differ in the acceptance condition applied to the run-tree $\tree$.
We carry on in this direction, and introduce a \emph{coloured} variant of the relational semantics,
designed in such a way that the tree automaton $\langle \, \Sigma \, , \, Q \, , \, \delta_f \, \rangle$
associated to a morphism $f: {!X}\tensor{!A}\to A$ defines a parity tree automaton.
This leads us to the definition of an inductive-coinductive fixpoint operator~$\fixpoint{}$
tightly connected to the current investigations on higher-order model-checking.

\paragraph*{Related works.}
The present paper is part of a wider research project devoted to the relationship 
between linear logic, denotational semantics and higher-order model-checking.
%
The idea developed here of shifting from the traditional finitary relational semantics
of linear logic to infinitary variants is far from new.
%
The closest to our work in this respect is probably the work by Miquel \cite{these-miquel}
where stable but non-continuous functions between coherence spaces are considered.
However, our motivations are different, since we focus here on the case 
of a modality $!A$ defined by finite-or-countable multisets in~$A$, 
which is indeed crucial for higher-order model-checking, but is not considered by Miquel.
%
In another closely related line of work, Carraro, Ehrhard and Salibra \cite{carraro-ehrhard-salibra}
formulate a general and possibly infinitary construction of the exponential modality~$A\mapsto {!A}$
in the relational model of linear logic.
However, the authors make the extra finiteness assumption in~\cite{carraro-ehrhard-salibra}
that the support of a possibly infinite multiset in $!A$ is necessarily finite.
Seen from that prospect, one purpose of our work is precisely to relax this finiteness condition which
appears to be too restrictive for our semantic account of higher-order model-checking
based on linear logic.
In a series of recent works,
Salvati and Walukiewicz \cite{salvati-walukiewicz2} \cite{salvati-walukiewicz3} have exhibited
a nice and promising connection between  higher-order model checking
and finite models of the simply-typed $\lambda$-calculus.
In particular, they establish the decidability of weak MSO properties of higher-order recursion schemes
by using purely semantic methods.
In comparison, we construct here a cartesian-closed category of sets and coloured relations
(rather than finite domains) where $\omega$-regular properties of higher-order recursion schemes
(and more generally of $\lambda\,Y$-terms) may be interpreted semantically thanks to a colour modality.
In a similar direction, Ong and Tsukada \cite{ong-tsukada} have recently constructed
a cartesian-closed category of infinitary games and strategies with similar connections
to higher-order model-checking.
Coming back to linear logic,
we would like to mention the works by Baelde \cite{baelde} and Montelatici \cite{montelatici}
who developed infinitary variants (either inductive-coinductive or recursive) of linear logic,
with an emphasis on the syntactic rather than semantic side.
In a recent paper working like we do here at the converging point of linear logic and automata theory, 
Terui \cite{terui} uses a qualitative variant of the relational semantics of linear logic
where formulas are interpreted as partial orders and proofs as downward sets in order to
establish a series of striking results on the complexity of normalization of simply-typed $\lambda$-terms.
Finally, an important related question which we leave untouched here is the comparison
between our work
and the categorical reconstruction of parity games achieved by Santocanale
\cite{santocanale2,santocanale} using the notion of bicomplete category,
see also his more recent work with Fortier \cite{santocanale3}.
%
%


\paragraph*{Plan of the paper.}
We start by recalling in \S\ref{section/rel} the traditional relational model of linear logic.
Then, after recalling in \S\ref{section/fixpoints} the definition of a Conway fixpoint operator in a Seely category,
we construct in \S\ref{section/finitary-fixpoint} such a Conway operator for the relational semantics.
We then introduce in \S\ref{section/infinitary-rel} our infinitary variant of the relational semantics,
and illustrate its expressive power in \S\ref{section/inductive-and-coinductive} by defining two
different Conway fixpoint operators.
Then, we define in \S\ref{section/coloured-modality} a coloured modality for the relational semantics,
and construct in \S\ref{section/y-colore} a Conway fixpoint operator in that framework.
We finally conclude in \S\ref{section/conclusion}.

\section{The relational model of linear logic}
\label{section/rel}
In order to be reasonably self-contained, we briefly recall the relational model of linear logic.
The category $\Rel$ is defined as the category with finite or countable sets as objects, and with binary relations between $A$ and $B$
as morphisms~$A\to B$.
The category $\Rel$ is symmetric monoidal closed, with tensor product defined as (set-theoretic) cartesian product,
and tensorial unit defined as singleton:
$$
\begin{tabular}{rclcrcl}
$A \tensor B$ &$\, = \,$ & $A \times B$
& \quad\quad\quad\quad\quad\quad & 
$1$ &$\, = \,$ & $\{\star\}$.
\end{tabular}
$$
%
Its internal hom (also called linear implication) $X \multimap Y$ simply defined as $X \tensor Y$.
%
%
Since the object $\bot\,=\,1\,=\,\{\star\}$ is dualizing,
the category $\Rel$ is moreover $\ast$-autonomous.
The category $\Rel$ has also finite products defined as
$$
\begin{tabular}{rcl}
$A \& B$ &$\quad = \quad$ & $\{(1,a) \ \vert \ a \in A\} \cup \{(2,b) \ \vert \ b \in B\} $
\end{tabular}
$$
with the empty set as terminal object $\top$.
As in any category with finite products, there is a diagonal morphism $\diag{A}:A \rightarrow A\, \&\, A$
for every object~$A$, defined
as
$$
\diag{A} \quad = \quad \{(a,\,(i,\,a))\ \vert\ i \in \{1,\,2\} \mbox{ and } a \in A\}
$$

Note that the category $\Rel$ has finite sums as well, since the negation $\negation{A}\ =\ A \multimap \bot$ 
of any object $A$ is isomorphic to the object $A$ itself.
%
All this makes $\Rel$ a model of multiplicative additive linear logic.
In order to establish that it defines a model of propositional linear logic, 
we find convenient to check that it satisfies the axioms
of a Seely category, as originally axiomatized by Seely \cite{seely} 
and then revisited by Bierman \cite{bierman}, see the survey \cite{models-of-linear-logic} for details.
To that purpose, recall that a \emph{finite multiset} over a set $A$ 
is a (set-theoretic) function $w\,:\,A \rightarrow \mathbb{N}$ with finite support,
where the support of $w$ is the set of elements of $A$ whose image is not equal to $0$.
%
%
The functor $!: \Rel\to \Rel$ is defined as

$$
\begin{tabular}{rcl}
$!\,A$ & $\quad = \quad$ & $\mathcal{M}_{fin} (A)$\\
$!\,f$ & $=$ & $\{([a_1, \cdots,\, a_n],\,[b_1,\cdots ,\, b_n]) \ \vert\
 \forall i,\,(a_i,\,b_i) \in f \}$\\
\end{tabular}
$$

\noindent
The comultiplication and counit of the comonad are defined as the digging and dereliction morphisms below:

$$
\begin{tabular}{rcl}
$\dig{A}$ & $\quad = \quad$ & $\{(w_1 + \cdots + w_k,\,[w_1, \cdots,\,w_k])\ \vert\ \forall i, \, w_i \in\ !\,A\} \ \in\ \Rel(!A,\,!!A)$\\
$\der{A}$ & $=$ & $\{([a],\,a) \ \vert\ a \in A \} \ \in\ \Rel(!A,\,A)$\\
\end{tabular}
$$

\noindent
In order to define a Seely category, one also needs the family of isomorphisms

$$
\begin{array}{ccccc}
\laxzero & \quad : \quad & 1 & \quad \longrightarrow \quad & ! \, \top
\\
\laxdeux{A}{B} & \quad : \quad & ! \, A \, \otimes  \,  ! \, B & \quad \longrightarrow \quad & ! \, (\, A \, \& \, B \, )
\end{array}
$$

\noindent
which are defined as $\laxzero = \{(\star,\,[])\}$ and

$$
\begin{tabular}{rcl}
$\laxdeux{A}{B}$ & $ = $ & $\{(([a_1, \cdots, a_m],[b_1,\cdots,b_n]),[(1,a_1), \cdots,\, (1,a_m),\,(2,b_1),\,\cdots,\,(2,b_n)])  \}$\\
\end{tabular}
$$

\noindent
One then carefully checks that the coherence diagrams expected of a Seely category commute.
From this follows that 

\begin{property}
The category $\Rel$ together with the finite multiset interpretation
of the exponential modality~$!$ defines a model of propositional linear logic.
\end{property}

\section{Fixpoint operators in models of linear logic}\label{section/fixpoints}
We want to extend linear logic with a fixpoint rule:

$$
\AxiomC{$!\,X \tensor \, !\, A \vdash A$}
\RightLabel{\quad $\fixrule$}
\UnaryInfC{$!\,X \vdash A$}
\DisplayProof
$$

\noindent
In order to interpret it in a Seely category, we need a parametrized fixpoint operator,
defined below as a family of functions

$$
\fixpoint{X,A}\ :\ \Ccategory(!\,X \, \tensor \, !A \, , \, A \,) \quad \morph{} \quad  \Ccategory(!\,X,A)
$$

\noindent
parametrized by $X, A$ and satisfying two elementary conditions,
mentioned for instance by Simpson and Plotkin in~\cite{simpson-plotkin}.
\begin{itemize}
\item \textbf{Naturality:} for any $g:{!\,X}\multimap Z$ and $f: {!\,Z} \, \tensor \, ! \, A \multimap A$, the diagram:

$$
\xymatrix@R=1em{
!\,X \ar[dd]_{\dig{X}} \ar[rrrr]^{\fixpoint{X,A}(k)} & & & & A\\
\\
!\,!\,X \ar[rrrr]_{!\,g} & & & & !\,Z \ar[uu]_{\ \ \fixpoint{Z,A}(f)}}
$$

\noindent
commutes, where the morphism $k:{!\,X} \tensor {!\,A} \multimap A$ in the upper part of the diagram is defined as the composite

$$
\xymatrix@R=1em{
!\,X \, \otimes \, ! \, A \ar[rrrr]^{k} \ar[dd]_{\dig{X} \,\tensor \, !A} & & & & A\\
\\
! \, ! \, X \, \otimes \, ! \, A \ar[rrrr]_{!\,g\,\tensor \, ! A} & & & & !\, Z\, \tensor \, ! \,A \ar[uu]_{f}\\
}
$$

\item \textbf{Parametrized fixpoint property:} for any $f: {!\,X} \,\tensor \, {!\,A} \multimap A$, the following diagram commutes:

$$
\xymatrix  {
!\,X \ar[d]_{!\,\diag{X}} \ar[rrrr]^{\fixpoint{X,A}(f)} & & & & A\\
!\,(\,X\,\&\,X\,) \ar[d]_{(\laxdeux{X}{X})^{-1}} & & &  & !\,X \, \tensor \, ! \, A \ar[u]_{f}\\
!\,X \, \otimes \, !\,X \ar[rrrr]_{!\,X\, \otimes\, \dig{X}} & & & &  !\,X \, \tensor \,!\,!\,X \ar[u]_{!\,X \, \tensor \, ! \, \fixpoint{X,A}(f)}\\ 
}
$$

\end{itemize}

These two equations are fundamental but they do not reflect all the equational properties 
of the fixpoint operator in domain theory.
For that reason, Bloom and Esik introduced the notion of \emph{Conway theory}
in their seminal work on iteration theories~\cite{bloom-esik,Bloom19961}.
This notion was then rediscovered and adapted to cartesian categories
by Hasegawa \cite{hasegawa},  by Hyland and by Simpson and Plotkin \cite{simpson-plotkin}.
Hasegawa and Hyland moreover independently established a nice correspondence
between the resulting notion of \emph{Conway fixpoint operator} and 
the notion of trace operator introduced a few years earlier
by Joyal, Street and Verity \cite{joyal-street-verity}.
Here, we adapt in the most straightforward way this notion of Conway fixpoint operator
to the specific setting of Seely categories.
Before going any further, we find useful to introduce the following notation: for every pair of morphisms

$$f\,:\,!\,X \,\tensor \,! \, B \multimap A \quad \mbox{  and  } \quad g\,:\,!\,X \,\tensor \, ! \, A \multimap B$$

\noindent
we write $f \star g\,:\,!\,X \, \otimes \, ! \, A  \multimap A$ for the composite:

$$
\makebox[10cm]{\xymatrix @R=2em{
!\,X \, \otimes \, ! \, A \ar[d]_{!\,\diag{X} \,\tensor\,!\,A} \ar[rrrr]^{f \star g} & & & & A\\
!\,(\,X \, \& \, X\,)\,\tensor\,!\,A \ar[d]_{(\laxdeux{X}{X})^{-1}\,\tensor\,!\,A} & & & & !\,X\, \tensor \, ! \, B \ar[u]_f\\
!\,X \, \tensor \, !\,X\,\tensor\,!\,A \ar[d]_{!\,X\,\tensor\,\laxdeux{X}{A}}  &&&& !\,X \, \tensor \, ! \, (\,!\,X\,\tensor\,!\,A\,) \ar[u]_{!\,X\,\tensor\,!\,g} \\
!\,X \, \tensor \, !\,(\,X\,\&\,A\,)  \ar[rrrr]_{!\,X\,\tensor\,\dig{X\& A}} &&   &&  !\,X \, \tensor \, ! \,!\,(\,X\,\&\,A\,) \ar[u]_{!\,X\,\tensor\,!\,(\laxdeux{X}{A})^{-1}} \\
}}
$$

\noindent
A Conway operator is then defined as a parametrized fixpoint operator satisfying the two additional properties below:
\begin{itemize}
\item \textbf{Parametrized dinaturality:} for any $f\,:\,!\,X \,\tensor \,! \, B \multimap A$ and $g\,:\,!\,X \,\tensor \, ! \, A \multimap B$, the following diagram commutes:
$$
\xymatrix @R=2em {
!\,X \ar[d]_{!\,\diag{X}} \ar[rrrr]^{\fixpoint{X,A}(f \star g)} & & & & A\\
!\,(\,X\,\&\,X\,) \ar[d]_{(\laxdeux{X}{X})^{-1}} & & &   & !\,X\,\tensor\,!\, B \ar[u]_f\\
!\,X\,\tensor\,!\,X \ar[rrrr]_{!\,X\,\tensor\,\dig{X}} &&&& !\,X \, \tensor \, !\,!\,X \ar[u]_{!\,X\,\tensor\, !\,\fixpoint{X,B}(g \star f)}\\
}
$$

\item \textbf{Diagonal property:} for every morphism $f\,:\,!\,X\,\tensor\,!\,A\,\tensor\,!\,A\,\multimap A$,

\begin{equation}
\label{eq/Y-diag}
\fixpoint{X,A}\,(\,(\laxdeux{X}{A})^{-1} \, \circ \,\fixpoint{X \& A,A}\,(\,f\, \circ\,(\,(\laxdeux{X}{A})^{-1} \, \tensor \, !\,A\,)\,)
\end{equation}
belongs to  $!\,X \multimap A$, since
$$
\xymatrix  {
!\,(\,X\,\&\,A \,) \,\tensor\,!\,A \ar[rr]^{(\laxdeux{X}{A})^{-1} \, \tensor \, !\,A} & & !\,X\,\tensor\,!\,A\,\tensor\,!\,A \ar[rr]^{f} & & A\\
}
$$

is sent by $\fixpoint{X \& A,A}$ to a morphism of $!\,(\,X \,\&\,A\,) \multimap A$, so that

$$
(\laxdeux{X}{A})^{-1} \, \circ \,\fixpoint{X \& A,A}\,(\,f\, \circ\,(\,(\laxdeux{X}{A})^{-1} \, \tensor \, !\,A\,)  \ : \ !\,X\,\tensor\,!\, A \multimap A
$$

to which the fixpoint operator $\fixpoint{X,A}$ can be applied, giving the morphism (\ref{eq/Y-diag}) of $!\,X \multimap A$. This morphism is required to
coincide with the morphism $\fixpoint{X,A}(k)$, where the morphism $k:{!X}\,\tensor\,{!A}\to A$ is defined as the composite
$$
\xymatrix@R=1em {
!\,X \, \tensor \,!\, A \ar[dd]_{!\,X\,\tensor\,!\,\diag{A}} \ar[rrrr]^{k} & & & & A \\
\\
!\,X\,\tensor\,!\,(\,A\,\&\,A\,) \ar[rrrr]_{!\,X\,\tensor\,(\laxdeux{A}{A})^{-1}} && & &!\,X \, \tensor \,!\, A \, \tensor \,!\, A \ar[uu]_f\\
}
$$
\end{itemize}
\noindent
Just as expected, we recover in that way the familiar notion of Conway fixpoint operator
as formulated in any cartesian category by Hasegawa, Hyland, Simpson and Plotkin:

\begin{property}
A Conway operator in a Seely category is the same thing as a Conway operator 
(in the sense of \cite{hasegawa,simpson-plotkin}) in the cartesian closed category
associated to the exponential modality by the Kleisli construction.
\end{property}
%
%

\section{A fixpoint operator in the relational semantics}
\label{section/finitary-fixpoint}

\red{
The relational model of linear logic can be equipped 
with a natural parameterized fixpoint operator $\fixpoint{}$ which transports any binary relation}
$$
\red{f\quad : \quad !\, X\,\tensor \, !\,A\, \quad \multimap \quad A}
$$
\red{to the binary relation}
$$
\red{\fixpoint{X,A}(f)\quad : \quad !\,X \multimap A}
$$
defined in the following way:
\begin{equation}\label{master-equation}
\begin{tabular}{rclr}
\hspace{-.8em}$\fixpoint{X,A}\,(f)$ & $\, = \,$ & $\{ \, (w,a) \, | \,$ &  $\exists \tree \in \runtree{f}{a} \,\, \mbox{with} \,\, w = \leaves{\tree} $\\
&&& $ \mbox{and } \tree \mbox{ is accepting} \, \}$\\
\end{tabular}
\end{equation}
where $\runtree{f}{a}$ is the set of ``run-trees'' defined as trees
with nodes labelled by elements of the set $X \uplus A$ and such that:
\begin{itemize}
\item the root of the tree is labelled by $a$,
\item the inner nodes are labelled by elements of the set $A$,
\item the leaves are labelled by elements of the set $X \uplus A$,
\item and for every node labelled by an element $b \in A$:
\begin{itemize}
\item if $b$ is an inner node, and letting $a_1, \cdots, a_n$ denote the labels 
of its children belonging to $A$ and $x_1, \cdots,\, x_m$ the labels belonging to $X$:
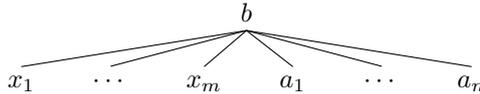
$$
\begin{tikzpicture}
\tikzset{level distance=25pt,sibling distance=15pt}
\Tree [.$b$ $x_1$ $\cdots$ $x_m$ $a_1$ $\cdots$ $a_n$ ]
\end{tikzpicture}
$$
then 
$
([(1,x_1),\cdots,\,(1,x_m),\,(2,a_1),\cdots ,\,(2,\,a_n)],b) \in f
$
\item if $b$ is a leaf, then $([],b) \in f$.
\end{itemize}
\end{itemize}
and where $\leaves{\tree}$ is the multiset obtained by enumerating the labels of the leaves of the run-tree $\tree$. 
Recall that multisets account for the number of occurences of an element, 
so that $\leaves{\tree}$ has the same number of elements as there are leaves
in the run-tree $\tree$.
Moreover, $\leaves{\tree}$ is independent of the enumeration of the leaves, 
since multisets can be understood as abelian versions of lists.
Finally, we declare that a run-tree is \emph{accepting} when it is a finite tree.
\begin{property}
The fixpoint operator $\fixpoint{}$ is a Conway operator on Rel.
\end{property}

\begin{example}
\label{example/witness}
Suppose that 
$$f\quad =\quad \{([],a)\} \cup \{([a,x],a)\}$$
where $A\,=\,\{a\}$ and $X\,=\,\{x\}$. Denote by $\mathcal{M}_n$ the finite multiset containing the element $x$ with multiplicity $n$. 
Then, for every $n \in \mathbb{N}$, we have that $(\mathcal{M}_n,a) \in \fixpoint{X,A}(f)$ 
since $(\mathcal{M}_n,a)$ can be obtained from the $\{a,\,x\}$-labelled witness run-tree 
of Figure \ref{figure/witness1}, which has $n+1$ internal occurrences of the element $a$, 
and $n$ occurrences of the element $x$ at the leaves. The witness tree is finite, so that it is accepted.
Now, consider the relation
$$g\quad =\quad \{([a],a)\} \cup \{([a,x],a)\}$$
In that case, $(\mathcal{M}_n,a)$ is not an element of $\fixpoint{X,A}(g)$ for any $n \in \mathbb{N}$
because all run-trees are necessarily infinite, as depicted in Figure \ref{figure/badwitness}, 
and thus, none is accepting. 
As a consequence, $\fixpoint{X,A}(g)$ is the empty relation.

\begin{figure}[t]
\begin{small}
\centering
\begin{minipage}{.5\textwidth}
  \centering
\begin{tikzpicture}
\Tree [.$a$ $x$ [.$a$ $x$ \edge[dotted]; [.$a$ $x$  [.$a$ ] ] ] ]
\end{tikzpicture}
  \captionof{figure}{An accepting run-tree.}
\label{figure/witness1}
\end{minipage}%
\begin{minipage}{.5\textwidth}
  \centering
 \begin{tikzpicture}
\tikzset{level distance=22pt}
\Tree [.$a$ $x$ [.$a$ $x$ \edge[dotted]; [.$a$ $x$ [.$a$ \edge[dotted]; $\ $ ] ] ] ]
\end{tikzpicture}
  \captionof{figure}{A non-accepting run-tree.}
\label{figure/badwitness}
\end{minipage}
\end{small}
\vspace{-0.5cm}
\end{figure}
\end{example}
%
The terminology which we have chosen for the definition of $\fixpoint{}$ is obviously automata-theoretic.
In fact, as we already mentioned in the introduction, this definition may be formulated
as an exploration of the infinitary tree $\textbf{comb}$ on the ranked alphabet $\Sigma\,=\,\{\,\bullet\,:\,2,\,\circ\,:\,0 \,\}$
by an alternating tree automaton associated to the binary relation $f\,:\,!\, X\,\tensor \, !\,A\, \multimap A$.
Indeed, given an element $a \in A$, 
consider the alternating tree automata $\mathcal{A}_{f,a}\,=\,\langle \Sigma,\, X \uplus A,\,\delta,\,a \rangle$ where, for $b \in A$ and $x \in X$:
$$
\delta(b,\,\bullet)= \bigvee_{(([x_1,\cdots,\,x_n),[a_1,\cdots,\,a_m]),b) \in f} \ \left(\,(1,x_1) \wedge \cdots \wedge (1,x_n) \wedge  (2,a_1) \wedge \cdots \wedge (2,a_m) \right)
$$
\begin{center}
\begin{tabular}{rcl}
$\delta(x,\,\bullet)\ =\ \bot$ &
$\quad \quad \delta(x,\,\circ)\ =\ \top \quad \quad $ &
$\delta(b,\,\circ)\ =\ \begin{cases} \top &\mbox{if } ([],b) \in f \\
\bot & \mbox{else} \end{cases}$\\
\end{tabular}
\end{center}
Note that we allow here the use of an infinite non-deterministic choice operator $\bigvee$ in formulas describing transitions, but only with \emph{finite} alternation.
Now, our point is that $\runtree{f}{a}$ coincides with the set of run-trees of the alternating automaton $\mathcal{A}_{f,a}$
over the infinite tree \textbf{comb} depicted in the Introduction. 
Notice that only finite run-trees are accepting: this requires that for some $b \in A$ the transition $\delta(b,\,\bullet)$ contains the alternating choice $\top$, in which the exploration of the infinite branch of \textbf{comb} stops and produces an accepting run-tree. This requires in particular the existence of some $b \in A$ such that $([],b) \in f$.

\section{Infinitary exponentials}
\label{section/infinitary-rel}
Now that we established a link with tree automata theory, it is tempting to relax the finiteness acceptance condition
on run-trees applied in the previous section.
%
To that purpose, however, we need to relax the usual assumption that the formulas of linear logic
are interpreted as finite or countable sets.
Suppose indeed that we want to interpret the exponential modality
$$\superbang A$$
as the set of finite or countable multisets, where a countable multiset
of elements of~$A$ is defined as a function
$$
A \quad \longrightarrow \quad \overline{\mathbb{N}}
$$
with finite or countable support.
%
Quite obviously, the set
$$
\superbang \, \mathbb{N}
$$
has the cardinality of the reals $\cardreals$.
We thus need to go beyond the traditionally countable relational interpretations of linear logic.
However, we may suppose that every set $A$ interpreting a formula has a cardinality 
below or equal $\cardreals$.
In order to understand why, it is useful to reformulate the elements of $\superbang A$
as finite or infinite words of elements of~$A$ modulo an appropriate notion of equivalence
of finite or infinite words up to permutation of letters.
Given a finite word $u$ and a finite or infinite word $w$, we write
$$
u \sqsubseteq w
$$
when there exists a finite prefix $v$ of $w$ such that $u$ is a prefix of $v$ modulo permutation of letter.
We write
$$
w_1 \simeq w_2 \stackrel{def}{\iff} \forall u \in A^{*}, \quad u\sqsubseteq w_1 \iff  u\sqsubseteq w_2
$$
where $A^{*}$ denotes the set of finite words on the alphabet~$A$.
\begin{proposition}
There is a one-to-one relationship between the elements of $\superbang A$
and the finite or infinite words on the alphabet $A$ modulo the equivalence relation~$\simeq$.
\end{proposition}
This means in particular that for every set~$A$,
there is a surjection from the set~$A^{\infty}=A^{\ast}\uplus A^{\omega}$
of finite or infinite words on the alphabet~$A$ to the set $\superbang A$
of finite or countable multisets.
An element of the equivalence class associated to a multiset is called a \emph{representation} of this multiset.
Notice that if a set~$A$ has cardinality at most $\cardreals$, the set~$A^{\infty}$
is itself bounded by $\cardreals$, since $(\cardreals)^{\aleph_0}\ =\ 2^{\aleph_0 \times \aleph_0}\ =\ \cardreals$.
This property leads us to define the following extension of $Rel$:


\begin{definition}
The category $\Relinfinitary$ has the sets $A,B$ of cardinality at most
$\cardreals$ as objects, and binary relations $f\subseteq A\times B$
between $A$ and $B$ as morphisms $A\to B$.
\end{definition}
Since a binary relation between two sets $A$ and $B$ is a subset of $A\times B$,
the cardinality of a binary relation in $\Relinfinitary$ is also bounded by $\cardreals$.
Note that the hom-set $\Relinfinitary(A,B)$ is in general of higher cardinality than $\cardreals$,
yet it is bounded by the cardinality of the powerset of the reals.
It is immediate to establish that:
\begin{property}
The category $\Relinfinitary$ is $\ast$-autonomous and has finite products.
As such, it provides a model of multiplicative additive linear logic.
\end{property}
There remains to show that the finite-or-countable multiset construction $\superbang$
defines a categorical interpretation of the exponential modality of linear logic.
Again, just as in the finitary case, we find convenient to check that $\Relinfinitary$ together 
with the finite-or-countable multiset interpretation $\superbang$ satisfy the axioms of a Seely category.
%
In that specific formulation of a model of linear logic, the first property to check is that:
\begin{property}
The finite-or-countable multiset construction $\superbang$ defines 
a comonad on the category $\Relinfinitary$.
\end{property}
The counit of the comonad is defined as the binary relation
$$
\der{A} \quad : \quad \superbang \, A \quad \longrightarrow \quad A
$$
which relates $[a]$ to $a$ for every element~$a$ of the set~$A$. 
In order to define its comultiplication, we need first to extend the notion of sum 
of multisets to the infinitary case, which we do in the obvious way, by extending
the binary sum of~$\mathbb{N}$ to possibly infinite sums in its completion $\overline{\mathbb{N}}$.
%
%
In order to unify the notation for finite-or-countable multisets with the one for finite multisets used in Section \ref{section/rel}, 
we find convenient to denote by $[a_1,\,a_2,\,\cdots]$ the countable multiset admitting the representation $a_1 a_2 \cdots$
We are now ready to describe the comultiplication 
$$\dig{A}\quad : \quad \superbang\, A \quad \rightarrow \quad \superbang\,\superbang\,A$$ 
of the comonad $\superbang$ as a straightforward generalization of the finite case:
$$
\begin{tabular}{rcl}
$\dig{A}$&$\quad = \quad $ &$\{(w_1 + \cdots + w_k,\,[w_1, \cdots,\,w_k])\ \vert\ \forall i \in \{1, \cdots n\}, \, w_i \in\ \superbang\,A\}$\\
&& $ \! \! \! \! \! \! \cup\ \{ (w_1 + \cdots + w_k + \cdots,\,[w_1, \cdots,\,w_k, \cdots])\ \vert\ \forall i \in \mathbb{N}, \, w_i \in\ \superbang\,A\}$\\
\end{tabular}
$$
One then defines the isomorphism
\begin{equation}\label{equation/iso-seely-0}
\laxzero \ = \ \{(\star,[])\} \quad : \quad 1 \quad \longrightarrow \quad \superbang \, \top
\end{equation}
and the family of isomorphisms
\begin{equation}\label{equation/iso-seely-2}
\laxdeux{A}{B} \quad : \quad
\superbang \, A \, \otimes  \, 
\superbang \, B
\quad \longrightarrow \quad 
\superbang \, (\, A \, \& \, B \, )
\end{equation}
indexed by the objects $A,B$ of the category~$\Relinfinitary$ which relates 
every pair $(w_A,w_B)$ of the set $\superbang \, A \, \otimes  \, \superbang \, B$ with the finite-or-countable multiset
$$
(\{1\}\times w_A) + (\{2\} \times w_B) \quad \in \quad \superbang \, (\, A \, \& \, B \, )
$$
where the operation $\{1\} \times w_A$ maps the finite-or-countable multiset $w_A\,=\,[a_1,\,a_2,\ldots]$ of elements of
$A$ to the finite-or-countable multiset  $[(1,\,a_1),\,(1,\,a_2),\ldots]$ of $\superbang (A \& B)$. We define
 $\{2\} \times w_B$ similarly.
We check carefully that
\begin{property}
The comonad $\superbang$ on the category $\Relinfinitary$ 
together with the isomorphisms (\ref{equation/iso-seely-0}) and (\ref{equation/iso-seely-2})
satisfy the coherence axioms of a Seely category -- see \cite{models-of-linear-logic}.
\end{property}

\noindent
In other words, this comonad $\superbang$ over the category $\Relinfinitary$ induces a new and infinitary model of propositional linear logic. 
The next section is devoted to the definition of two different fixpoint operators living inside this new model.

\section{Inductive and coinductive fixpoint operators}\label{section/inductive-and-coinductive}
In the infinitary relational semantics, a binary relation 
$$f\,\,:\,\,\superbang A \, \, \multimap \, \, B$$
may require a countable multiset~$w$ of elements (or positions) of the input set~$A$
in order to reach a position~$b$ of the output set~$B$. 
For that reason, we need to generalize the notion of alternating tree automata to \emph{finite-or-countable alternating tree automata}, 
a variant in which formulas defining transitions use of a possibly countable alternation operator $\bigwedge$ 
and of a possibly countable non-deterministic choice operator $\bigvee$.
The generalization of the family of automata $\mathcal{A}_{f,a}$ of \S\ref{section/finitary-fixpoint} leads to a new definition of the set $\runtree{f}{a}$, in which witness trees may have internal nodes of countable arity.
A first important observation is the following result:

\begin{property}
Given $f\,:\,\superbang\, A \, \otimes \, \superbang \,X \multimap A$, $a \in A$, and $\tree \in \runtree{f}{a}$, the multiset $\leaves{\tree}$ is finite or countable.
\end{property}
An important consequence of this observation is that the definition of the Conway operator $\fixpoint{}$ given in Equation~(\ref{master-equation}) 
can be very simply adapted to the finite-or-countable interpretation of the exponential modality~$\superbang$ in the Seely category $\Relinfinitary$.
Moreover, in this infinitary model of linear logic, we can give more elaborate acceptation conditions, among which two are canonical:
\begin{itemize}
\item considering that any run-tree is accepting, one defines the \emph{coinductive} fixpoint on the model, which is the greatest fixpoint over $\Relinfinitary$.
\item on the other hand, by accepting only trees without infinite branches, we obtain the \emph{inductive} interpretation of the fixpoint,
which is the least fixpoint operator over $\Relinfinitary$.
\end{itemize}

It is easy to see that the two fixpoint operators are different: recall Example~\ref{example/witness},
and observe that the binary relation~$g$ is also a relation in the infinitary semantics.
It turns out that its inductive fixpoint is the empty relation, while its coinductive fixpoint coincides with the relation
$$
\{\mathcal{M}_n,a) \ \vert \ \forall n \in \mathbb{N}\}\ \cup\ \{([x,\,x,\,\cdots],a)\}
$$
In this coinductive interpretation, the run-tree obtained by using infinitely $([x,a],a)$ and never $([a],a)$ is accepting and is the witness tree generating $\{([x,\,x,\,\cdots],a)\}$.

\begin{property}
The inductive and coinductive fixpoint operators over the infinitary relational model of linear logic are Conway operators on this Seely category.
\end{property}

\section{The coloured exponential modality}\label{section/coloured-modality}
In their semantic study of the parity conditions used in higher-order model-checking,
and more specifically in the work by Kobayashi and Ong \cite{kobayashi-ong},
the authors have recently discovered~\cite{coloured-tensorial-logic} that these parity conditions are secretly regulated
by the existence of a comonad $\modality$ which can be interpreted in the relational semantics of linear logic
as 
$$
\modality\ A \quad = \quad Col \times A
$$
where $Col=\{1,\dots,N\}$ is a finite set of integers called \emph{colours}.
The colours (or priorities) are introduced in order to regulate the fixpoint discipline:
in the immediate scope of an even colour, fixpoints should be interpreted coinductively,
and inductively in the immediate scope of an odd colour.
It is worth mentioning that the comonad~$\modality$ has its comultiplication defined by the maximum operator
in order to track the maximum colour encountered during a computation:
$$
\begin{array}{ccccccc}
\delta_A &\, = \, &\{ (max(c_1,c_2), a) , (c_1, (c_2, a))) \, | \, c_1,c_2\in Col, a\in A\} &\hspace{.5em} : \hspace{.5em}&
\modality A & \hspace{.5em} \multimap  \hspace{.5em} & \modality \, \modality A\\
\varepsilon_A & =  & \{ (1,a) , a) \, | \, a\in A\} &  :  & \modality A &  \multimap  & A
\end{array}
$$
whereas the counit is defined using the minimum colour $1$.
%
The resulting comonad is symmetric monoidal and also satisfies the following key property:

\begin{property}
There exists a distributive law $\lambda \, : \,  \superbang \ \modality \,\, \rightarrow \,\, \modality \ \superbang$ between comonads.
\end{property}

\noindent
A fundamental consequence is that the two comonads can be composed 
into a single comonad $\colorbang$ defined as follows:
$$
\colorbang \quad  = \quad  \superbang \ \circ \ \modality
$$
The resulting \emph{infinitary} and \emph{coloured} relational semantics of linear logic
is obtained from the category $\Relinfinitary$ equipped with the composite comonad $\colorbang$.

\begin{theorem}
The category $\Relinfinitary$ together with the comonad $\colorbang$
defines a Seely category and thus a model of propositional linear logic.
\end{theorem}

\section{The inductive-coinductive fixpoint operator~$\fixpoint{}$}
\label{section/y-colore}
We combine the results of the previous sections
in order to define a fixpoint operator $\fixpoint{}$ over the infinitary coloured relational model, 
which generalizes both the inductive and the coinductive fixpoint operators.
%
Note that in this infinitary and coloured framework,
we wish to define a fixpoint operator $\fixpoint{}$ which transports a binary relation
$$
f\quad : \quad\colorbang X \, \otimes \, \colorbang A \,\,\, \multimap \,\,\, A
$$
into a binary relation
$$
\fixpoint{X,A}\,(f) \quad : \quad \colorbang X \,\,\, \multimap \,\,\, A.
$$
To that purpose, notice that the definition given in \S\ref{section/finitary-fixpoint}
of the set $\runtree{f}{a}$ of run-trees
extends immediately to this new coloured setting, since the only change is in the set of labellings. 
Again, accepting all run-trees would lead to the coinductive fixpoint, while accepting only run-trees whose branches are finite 
would lead to the inductive fixpoint. 
We now define our acceptance condition for run-trees in the expected way, 
directly inspired by the notion of alternating parity tree automaton.
Consider a run-tree $\tree$, and remark that its nodes are labelled with elements 
of $(\,Col \times A \,) \cup (\,Col \times X\,)$. 
We call the colour of a node the first element of its label. Coloured acceptance is then defined as follows:
\begin{itemize}
\item a finite branch is accepting,
\item an infinite branch is accepting precisely when the greatest colour appearing infinitely often in the labels of its nodes is even.
\item a run-tree is accepting precisely when all its branches are accepting.
\end{itemize}
Note that a run-tree whose nodes are all of an even colour will be accepted independently of its depth, 
as in the coinductive interpretation, while a run-tree labelled only with odd colours will be accepted 
precisely when it is finite, just as in the inductive interpretation. 
We call the fixpoint operator associated with the notion of coloured acceptation 
the inductive-coinductive fixpoint operator over the infinitary coloured relational model.

\begin{theorem}
The inductive-coinductive fixpoint operator $\fixpoint{}$ defined over the infinitary coloured relational semantics of linear logic
is a Conway operator.
\end{theorem}

\section{Conclusion}\label{section/conclusion}

In this article, we introduced an infinitary variant of the familiar relational semantics of linear logic.
We then established that this infinitary model accomodates an inductive as well as a coinductive Conway operator $\fixpoint{}$.
This propelled us to define a coloured relational semantics and to define an inductive-coinductive fixpoint operator
based on a parity acceptance condition.
%
%
The authors proved recently \cite{coloured-tensorial-logic} that a recursion scheme can be interpreted in this model in such a way 
that its denotation contains the initial state of an alternating parity automaton if and only if the tree 
it produces satisifies the MSO property associated to the automaton. 
A crucial point related to the work by Salvati and Walukiewicz \cite{salvati-walukiewicz1}
is the fact that a tree satisfies a given MSO property if and only if any suitable representation 
as an infinite tree of a $\lambda Y$-term generating it also does.
We are thus convinced that this infinitary and coloured variant of the relational semantics of linear logic
will play an important and clarifying role in the denotational and compositional study of higher-order model-checking.

\bibliography{fossacs}

\begin{thebibliography}{10}

\bibitem{baelde}
David Baelde.
\newblock Least and greatest fixed points in linear logic.
\newblock {\em {ACM} Trans. Comput. Log.}, 13(1):2, 2012.

\bibitem{bierman}
Gavin~M. Bierman.
\newblock What is a categorical model of intuitionistic linear logic?
\newblock In Mariangiola Dezani{-}Ciancaglini and Gordon~D. Plotkin, editors,
  {\em {TLCA} '95, Edinburgh, UK, April 10-12, 1995, Proceedings.}, volume 902
  of {\em LNCS}, pages 78--93. Springer, 1995.

\bibitem{bloom-esik}
S.L. Bloom and Z.~{\'E}sik.
\newblock {\em Iteration theories: the equational logic of iterative
  processes}.
\newblock EATCS monographs on theoretical computer science. Springer-Verlag,
  1993.

\bibitem{Bloom19961}
Stephen~L. Bloom and Zoltán Ésik.
\newblock Fixed-point operations on ccc's. part i.
\newblock {\em Theoretical Computer Science}, 155(1):1 -- 38, 1996.

\bibitem{carraro-ehrhard-salibra}
Alberto Carraro, Thomas Ehrhard, and Antonino Salibra.
\newblock Exponentials with infinite multiplicities.
\newblock In Anuj Dawar and Helmut Veith, editors, {\em {CSL} 2010, Brno, Czech
  Republic, August 23-27, 2010. Proceedings.}, volume 6247 of {\em LNCS}, pages
  170--184. Springer, 2010.

\bibitem{santocanale3}
J{\'e}r{\^o}me Fortier and Luigi Santocanale.
\newblock Cuts for circular proofs: semantics and cut-elimination.
\newblock In Simona Ronchi~Della Rocca, editor, {\em CSL}, volume~23 of {\em
  LIPIcs}, pages 248--262. Schloss Dagstuhl - Leibniz-Zentrum fuer Informatik,
  2013.

\bibitem{coloured-tensorial-logic}
Charles Grellois and Paul{-}Andr{\'{e}} Melli{\`{e}}s.
\newblock Tensorial logic with colours and higher-order model checking.
\newblock submitted, 2015.

\bibitem{hasegawa}
Masahito Hasegawa.
\newblock {\em Models of Sharing Graphs: A Categorical Semantics of Let and
  Letrec}.
\newblock Number 1192 in Distinguished dissertations series. Springer-Verlag,
  1999.

\bibitem{joyal-street-verity}
Andr{\'{e}} Joyal, Ross Street, and Dominic Verity.
\newblock Traced monoidal categories.
\newblock {\em Mathematical Proceedings of the Cambridge Philosophical
  Society}, 119:447--468, 4 1996.

\bibitem{kobayashi-ong}
Naoki Kobayashi and C.-H.~Luke Ong.
\newblock A type system equivalent to the modal mu-calculus model checking of
  higher-order recursion schemes.
\newblock In {\em LICS}, pages 179--188. IEEE Computer Society, 2009.

\bibitem{models-of-linear-logic}
Paul{-}Andr{\'{e}} Melli{\`{e}}s.
\newblock Categorical semantics of linear logic.
\newblock In {\em Interactive models of computation and program behaviour},
  pages 1--196. 2009.

\bibitem{these-miquel}
Alexandre Miquel.
\newblock {\em Le calcul des constructions implicites : syntaxe et
  s{\'{e}}mantique}.
\newblock PhD thesis, Universit{\'{e}} Paris 7, 2001.

\bibitem{montelatici}
Rapha{\"{e}}l Montelatici.
\newblock Polarized proof nets with cycles and fixpoints semantics.
\newblock In {\em {TLCA} 2003 Proceedings.}, 2003.

\bibitem{salvati-walukiewicz1}
Sylvain Salvati and Igor Walukiewicz.
\newblock Evaluation is msol-compatible.
\newblock In Anil Seth and Nisheeth~K. Vishnoi, editors, {\em {FSTTCS}},
  volume~24 of {\em LIPIcs}, pages 103--114. Schloss Dagstuhl - Leibniz-Zentrum
  fuer Informatik, 2013.

\bibitem{salvati-walukiewicz2}
Sylvain Salvati and Igor Walukiewicz.
\newblock Using models to model-check recursive schemes.
\newblock In Masahito Hasegawa, editor, {\em {TLCA} 2013, Eindhoven, The
  Netherlands, June 26-28, 2013. Proceedings.}, volume 7941 of {\em LNCS},
  pages 189--204. Springer, 2013.

\bibitem{salvati-walukiewicz3}
Sylvain Salvati and Igor Walukiewicz.
\newblock {Typing weak MSOL properties}.
\newblock September 2014.

\bibitem{santocanale2}
Luigi Santocanale.
\newblock A calculus of circular proofs and its categorical semantics.
\newblock In Mogens Nielsen and Uffe Engberg, editors, {\em FoSSaCS}, volume
  2303 of {\em Lecture Notes in Computer Science}, pages 357--371. Springer,
  2002.

\bibitem{santocanale}
Luigi Santocanale.
\newblock $\mu$-bicomplete categories and parity games.
\newblock {\em ITA}, 36(2):195--227, 2002.

\bibitem{seely}
R.A.G. Seely.
\newblock Linear logic, -autonomous categories and cofree coalgebras.
\newblock In {\em In Categories in Computer Science and Logic}, pages 371--382.
  American Mathematical Society, 1989.

\bibitem{simpson-plotkin}
Alex~K. Simpson and Gordon~D. Plotkin.
\newblock Complete axioms for categorical fixed-point operators.
\newblock In {\em {LICS} 2000, USA, June 26-29, 2000}, pages 30--41. {IEEE}
  Computer Society, 2000.

\bibitem{terui}
Kazushige Terui.
\newblock Semantic evaluation, intersection types and complexity of simply
  typed lambda calculus.
\newblock In Ashish Tiwari, editor, {\em RTA}, volume~15 of {\em LIPIcs}, pages
  323--338. Schloss Dagstuhl - Leibniz-Zentrum fuer Informatik, 2012.

\bibitem{ong-tsukada}
Takeshi Tsukada and C.{-}H.~Luke Ong.
\newblock Compositional higher-order model checking via
  \emph{{\(\omega\)}}-regular games over b{\"{o}}hm trees.
\newblock In Thomas~A. Henzinger and Dale Miller, editors, {\em {CSL-LICS} '14,
  Vienna, Austria, July 14 - 18, 2014}, page~78. {ACM}, 2014.

\end{thebibliography}
\bibliographystyle{plain}

\section*{Appendix: Seely categories}
A \emph{Seely category} is defined as a symmetric monoidal closed category
$(\LAT,\tensor,1)$ with binary products $A\with B$, a terminal object~$\top$, and:
\begin{enumerate}
\item a comonad $(!,\dig{},\der{})$,
\item two natural isomorphisms 
\begin{center}
\begin{tabular}{ccc}
$\laxdeux{A}{B}
\hspace{1em}
:
\hspace{1em}
!A\tensor !B
\hspace{.3em}
\cong 
\hspace{.3em}
!(A\with B)$
&
\hspace{4em}
&
$\laxzero
\hspace{1em}
:
\hspace{1em}
1
\hspace{.3em}
\cong 
\hspace{.3em}
!\top$
\end{tabular}
\end{center}
making
$$(!,\lax) \quad : \quad (\LAT,\with,\top) \quad \morph{} \quad (\LAT,\tensor,\one)$$
a symmetric monoidal functor.
\end{enumerate}
One also asks that the coherence diagram
\begin{equation}
\label{equation/diagram-of-seely-one}
\vcenter{\xymatrix @-.5pc {
!A\tensor !B
\ar[rrrr]^{\lax}
\ar[dd]_{\dig{A}\tensor \dig{B}}
&&&&
!(A\with B)
\ar[d]^{\dig{A\with B}}
\\
&&&&
!!(A\with B)
\ar[d]^{!\paire{!\pi_1}{!\pi_2}}
\\
!!A\tensor !!B
\ar[rrrr]^{\lax}
&&&&
!(!A\with !B)}}
\end{equation}
commutes in the category~$\LAT$ for all objects~$A$ and~$B$, and that the four following diagrams expressing the fact that the functor~$(!,\lax)$
is symmetric monoidal:
\begin{equation}\label{equation/diagram-of-seely-two}
\vcenter{\xymatrix @+.2pc {
(!A\tensor !B)\tensor !C
\ar[rrr]^{\alpha}
\ar[d]_{\lax\tensor !C}
&&&
!A\tensor(!B\tensor !C)
\ar[d]^{!A\tensor \lax}
\\
!(A\with B)\tensor !C\ar[d]_{\lax}
&&&
!A\tensor !(B\with C)\ar[d]^{\lax}
\\
!((A\with B)\with C)\ar[rrr]^{!\alpha}
&&&
!(A\with (B\with C))}}
\end{equation}
\begin{equation}\label{equation/diagram-of-seely-three}
\begin{array}{ccc}
\vcenter{\xymatrix @-.1pc {
!A\tensor 1
\ar[rr]^{\rho}
\ar[d]_{!A\tensor \lax}
&&
!A\\
!A\tensor !\top\ar[rr]^{\lax}
&&
!(A\with\top)\ar[u]_{!\rho}}}
&&
\vcenter{\xymatrix @-.1pc {
1\tensor !B\ar[rr]^{\lambda}\ar[d]_{\lax\tensor !B} && !B\\
!\top\tensor !B\ar[rr]^{\lax}&&!(\top\with B)\ar[u]_{!\lambda}}}
\end{array}
\end{equation}
\begin{equation}\label{equation/diagram-of-seely-four}
\vcenter{
\xymatrix @+.6pc {
  !A\tensor !B\ar[rr]^{\gamma} \ar[d]_{\lax}
&& 
  !B\tensor !A\ar[d]^{\lax}
\\
  !(A\with B)\ar[rr]^{!\gamma}
&& 
  !(B\with A)
}}
\end{equation}
commute in the category~$\LAT$ for all objects~$A,B$ and $C$.

\end{document}